# Enabling the autofocus approach for parameter optimization in planar measurement geometry clinical optoacoustic imaging

Ludwig Englert[1,2], Lucas Riobo[1,2], Christine Schonemann[1,2], Vasilis Ntziachristos[1,2,*], Juan Aguirre[1,2,*]

[1] Institute for Biological and Medical Imaging (IBMI), Helmholtz Zentrum München, D-85764 Neuherberg, Germany.
[2] Chair for Biological Imaging, TranslaTUM, Technische Universität München (TUM), D-81675 München, Germany.

∗ Corresponding authors: Juan Aguirre (juanaguir@gmail.com) and Vasilis Ntziachristos (bioimaging.translatum@tum.es).

## Key words

## Abstract

In optoacoustic (photoacoustic) tomography, several parameters related to tissue and detector features are needed for image formation, but they may not be known *a priori*. An autofocus (AF) algorithm is generally used to estimate these parameters. However, the algorithm works iteratively, therefore, it is impractical for clinical imaging with systems featuring planar geometry due to long reconstruction times.

We have developed a fast autofocus (FAF) algorithm for optoacoustic systems with planar geometry that is much simpler computationally than the conventional AF algorithm. We show that the FAF algorithm required about 5 sec. to provide accurate estimates of the speed of sound in simulated data and experimental data obtained using an imaging system that is poised to enter the clinic. The applicability of FAF for estimating other image formation parameters is discussed.

We expect the FAF algorithm to contribute decisively to the clinical use of optoacoustic tomography systems with planar geometry.

# Introduction

Optoacoustic (photoacoustic) tomography is an emerging optical imaging modality that combines the rich contrast based on optical absorption with the ability of ultrasound to provide high-resolution images deep inside tissue [1-3]. In this technique, pulsed laser light is selectively absorbed by chromophores, creating a local increase in pressure via thermo-elastic expansion, which in turn produces ultrasound waves. Several detectors at the tissue surface record the acoustic waves over time. By applying tomographic principles, an image of the initial pressure distribution, and thus the regions of high optical absorption, can be obtained.

Optoacoustic tomography has taken a major step toward clinical use with the advent of imaging systems based on planar geometry [4-6]. In this geometry, the tissue is illuminated with unfocused pulsed light, and ultrasound detectors are arranged on a 2D plane above the tissue. Applying tomographic principles generates a 3D image of the tissue contained within the cube immediately below the 2D plane. The light emitter and detectors can be placed on the same side of the imaged region, creating a reflection-like configuration that can image regions that would be inaccessible to systems based on transmission or on cylindrical or spherical geometries. In a particularly promising approach, one focused ultra-wideband transducer is raster-scanned on a 2D plane above the surface of the tissue. This approach, known as Raster Scanning Optoacoustic Mesoscopy (RSOM), can achieve resolutions of tens of microns at depths of several millimetres and has demonstrated unique abilities for imaging tissue pathology in patients with dermatological disease, diabetes or cardiovascular disorders [7-10].

In RSOM, the ultrasound detector is scanned along a "fast" axis and a "slow" axis (Fig. 1a). At each point, 1D time domain ultrasound signals within the cone-shaped sensitivity field of the focused detector are collected, making up the so-called "A-scan". One-dimensional scanning of the detector along the fast axis generates a two-dimensional sinogram called the "B-scan", while scanning along the slow axis generates a stack of B-scans that together comprise a 3D sinogram called the "C-scan".

High-quality image formation algorithms for clinical applications requires knowledge of several parameters related to the acoustic properties of the tissue being imaged, such as the speed of sound. The speed of sound can vary between experiments, since it depends on the tissue composition as well as temperature [11]. Other parameters important for imaging relate to the detector, such as its numerical aperture or focal distance. Some of these parameters may not be known with sufficient accuracy, requiring their estimation.

In the simplest approach to estimating image formation parameters, values are defined manually and the resulting reconstructed images are visually inspected. This trial-and-error approach takes substantial time and reduces reproducibility, leading to the use of the autofocus (AF) algorithm[12] in optoacoustic tomography [11, 13].

The AF algorithm works by automatically reconstructing images covering a range of possible values for the particular imaging parameter; for example, it generates images in which the speed of sound ranges from 1,400 to 1,600 m/s in steps of 1 m/s. For each parameter value, a sharpness metric is calculated for the resulting image, forming a cost function. Appropriate selection of the sharpness metric results in a healthy cost function with a Gaussian or Lorentzian shape, whose maximum can be obtained easily using brute force. The parameter value yielding the maximum is taken as the correct estimate.

The AF algorithm has been validated for estimating speed of sound in imaging systems with cylindrical or planar geometries, but it requires extensive time when geometry is planar[11]. In a cylindrical configuration, 256 detectors are used, leading to fast reconstructions (<< 1 sec), which translates to approximately 1 sec to calculate the 50 reconstructions typically needed for the AF algorithm [13]. In some planar configurations, in contrast, the large number of detectors (~$10^4$) results in excessive computational time[11]. The time is even longer in RSOM, which may involve up to $10^5$ detector positions. Common implementations of time-domain back-projection reconstruction[14], which involves a complexity of $O(n^5)$, generally take around 2 min to perform one reconstruction and therefore around 1.5 h to calculate the AF algorithm. Even frequency-domain backprojection[14], which has a lower computational complexity of $O(n^3 log_2(n))$, requires approximately 15 sec per reconstruction[15] and therefore around 15 min to calculate the AF algorithm. Reconstruction times are even longer for other image formation approaches, such as model-based reconstruction[16, 17].

These computation times are far too long to support clinical applications, where far less than 1 min is available for calculating the AF algorithm. In order to fulfil this "< 1 min criterion" and move optoacoustic imaging with planar geometry closer to the clinic, we developed a "fast AF" (FAF) algorithm. The FAF leverages the fact that due to the symmetries in planar geometry, the B-scan signal from a 3D distribution of optoacoustic emitters can be approximated as the B-scan signal generated by a virtual 2D distribution of sources obtained by projecting the real 3D sources onto a "B-plane" associated with that B-scan (see Methods and Fig. 1a). As a result, the AF algorithm can be correctly applied to single B-scans by calculating a sharpness metric for the 2D reconstructions of the virtual sources (see Methods and Fig. 1a). This process substantially reduces the computational complexity of image formation from $O(n^5)$ to $O(n^3)$ in the case of time-domain reconstructions, and from $O(n^3 log_2(n))$ to $O(n^2 log_2(n))$ in the case of frequency-domain reconstructions[14] (also known as w-k algorithms[15]). This reduction in complexity translates to faster computation of the FAF algorithm.

In this paper we validate the FAF algorithm for estimating speed of sound in simulations, phantoms and clinical data collected using a standard RSOM system. The FAF algorithm was able to estimate speed of sound in clinical data within 7 sec. We compare the FAF algorithm against the conventional AF applied to 3D data, and we discuss the possibility of using the FAF algorithm to estimate other image formation parameters.

## **Materials and Methods**

### **Description of the FAF algorithm for estimating speed of sound**

For calculating the FAF algorithm, a complete C-scan is acquired, then a B-scan is randomly selected from the 3D sinogram. Several 2D reconstructions on the B-plane (associated with the B-scan) are performed, each with a different speed of sound. Next, a metric is used to assess the sharpness of the resulting 2D images or the 1D maximum intensity projection (MIP) of the 2D images along the z dimension (see "performance assessment of the FAF algorithm"). In this way, the distribution of sharpness as a function of the speed of sound is obtained. The estimation of the speed of sound corresponds to the value that leads to the maximum of the sharpness function.

The standard AF algorithm is based on 3D reconstructions of C-scans or the 2D MIPs of those 3D reconstructions. We hypothesized that applying the FAF algorithm to 2D reconstructions of B-scans or 1D MIPs of those 2D reconstructions would lead to the same estimated speed of sound as the conventional AF algorithm.

## **Justification of FAF assumptions**

In this section we proof that the B-scan from any 3D distribution of sources is equivalent to the signal from a specific 2D distribution of virtual sources on the B-plane associated with that B-scan (see Fig. 1a). Since the 2D reconstruction from such a B-scan displays the virtual sources, sharpness metrics can be applied to estimate the speed of sound, exactly as when operating with 3D reconstruction of real sources.

The proof relies on two assumptions. First, the optoacoustic signal from any 3D light-absorbing structure is equivalent to the signal generated by assuming that the absorbing structure is composed of small, discrete absorbing spheres. Second, the optoacoustic signal from a real sphere in the 3D space recorded on a B-scan is equivalent to the signal from a virtual sphere generated by projecting the real sphere onto the B-plane associated with the B-scan. This projection is performed along the angular dimension of a cylindrical coordinate system whose z axis aligns with the B-scan (see Fig. 1a).

Mathematically the two assumptions can be expressed as follows:

$$p(\vec{r_i}, t) \approx \sum_{j=1}^{N} p_o(\vec{r_j}) \, p_{sp}(\vec{r_i}, \vec{r_j}, t) \approx \sum_{j=1}^{N} p_o(\vec{r_j}') \, p_{sp}(\vec{r_i}, \vec{r_j}', t) \quad (1)$$

where $\vec{r_i}$ is the position of the l-th detector of a specific B-scan, $p(\vec{r_i}, t)$ is the optoacoustic signal arriving at the l-th detector at time $t$, $N$ is the total number of discrete spheres that make up the light-absorbing structure, $p_o(\vec{r_j})$ is the initial optoacoustic pressure ($t = 0$) inside the j-th sphere located at $\vec{r_j}$, $p_{sp}(\vec{r_i}, \vec{r_j}, t)$ is the optoacoustic pressure at time t inside the j-th sphere assuming an initial pressure of $1 \, N/m^2$ and measured at the l-th detector, and $p_{sp}(\vec{r_i}, \vec{r_j}', t)$ is the optoacoustic pressure in the j-th sphere after projection onto the B-plane at location $\vec{r_j}'$ (see Fig. 1a). We note that $p_o(\vec{r_j}') = p_o(\vec{r_j})$.

The first assumption in equation 1 holds as long as the spheres are sufficiently small, according to Huygens principle. In order to empirically validate the second assumption in equation 1, we simulated a B-scan corresponding to the 3D distribution of absorbers randomly distributed inside a volume. We then projected the absorbers onto the B-plane associated with the B-scan and calculated the corresponding "virtual" B-scan. Finally, we confirmed that subtraction of the B-scans from each other resulted in a blank image (Fig. 1b-d).

The second assumption in equation 1 can also be validated analytically if one notices that $p_{sp}(\vec{r_i}, \vec{r_j}, t)$ depends on the distance from the source and the detector [18], and that the angle between *r* and *z* is always 90º (Fig. 1a).

# Assessment of the FAF algorithm

## RSOM system and image reconstruction

To assess the performance of the FAF algorithm, we used an RSOM system developed by our group. The system features a custom-made, spherically focused piezoelectric 50-MHz transducer (bandwidth 10-120 MHz) with 3-mm focal distance (Sonaxis, Besancon, France). Signals were digitized using a 700-MHz data acquisition card working at 1 Gs per sec (GageScope RazorMAX; GaGe, Lockport, IL, USA). Optoacoustic signals were induced using a 1-nsec DPSS laser operating at 532 nm and a pulse rate of 500 Hz (Wedge HB; Bright Solutions, Cura Carpignano, Italy). The scan head with the transducer was mounted at the end of an articulated arm. It provided three motorized stages to allow precise positioning of the transducer in the x, y and z directions; the head was scanned in the x and y directions. Laser light was delivered below the transducer via a custom-made fiber bundle (Ceramoptec, Livani, Lituane). The entire system was controlled using MATLAB on a PC running Microsoft Windows. Further details of the system can be found elsewhere [7, 9].

All 2D reconstructions of B-scans as well as 3D reconstructions were performed using the 2D and 3D Fourier domain reconstructions in the k-wave toolbox in MATLAB [19]. The FAF algorithm, including calculation and evaluation of the sharpness metrics (see below), was coded entirely using MATLAB.

## Performance assessment

We assessed the performance of the FAF algorithm using simulations, phantoms, and clinical data.

Simulations:

We simulated single B-scans corresponding to 100, 3D distributions of 100 spherical absorbers with diameters between 10 and 30 µm, randomly distributed in a volume of x = 1800 µm, y = 200 µm and z = 1800 µm. The speed of sound was set to 1,550 m/s. For each of the 100 simulations the FAF algorithm was applied on a B-scan corresponding to a B-plane of 1800 x 1800 µm located at the center of the volume. More in detail, for each B-scan we calculated 40 reconstructions (speed of sound values ranging from 1,450 to 1,650 m/s in steps of 5 m/s). For each set of 40 reconstructions four speed of sound estimations were obtained from the cost functions that results from applying the four sharpness metrics described in "sharpness metrics". The most suitable sharpness metric should provide the clearest Gaussian/Lorentzian shape and the speed of sound estimation closest to the value of the forward solution.

Phantoms:

A 10-µm polyamide suture was immersed in water and imaged. Data were acquired in a scan window measuring 4 by 3 mm, with a step size of 15 µm in the x and y directions. This generated 36,180 A-lines. The FAF algorithm was applied to 10 equidistant B-scans along the slow axis at a step size of 30 µm in order to assess the stability of the estimated speed of sound along the field of view. Such stability is taken as an indicator of the performance of the algorithm. For each B-scan, we performed 29 reconstructions which correspond to speed of sound values varying from 1,420 to 1,560 m/s in steps of 5 m/s. For each set of 29 reconstructions four speed of sound estimations were obtained from the cost functions that results from applying the four sharpness metrics described in "sharpness metrics". The most suitable sharpness metric should provide the clearest Gaussian/Lorentzian shape and the speed of sound value most invariant to B-scan selection.

Clinical data:

An eczematous skin lesion was scanned to yield 54,135 A-lines over a window measuring 6 by 2 mm in steps of 15 µm in the x and y directions. The FAF algorithm was applied to 10 equidistant B-scans along the slow axis at a step size of 30 µm in order to investigate how stable the estimated speed of sound was along the field of view. Such stability is taken as an indicator of the performance of the algorithm. More in detail, we selected speed of sound values ranging from 1,440 to 1,620 m/s in steps of 5 m/s leading to 36 reconstructions for each B-scan. For each set of 36 reconstructions four speed of sound estimations were obtained from the cost functions that results from applying the four sharpness metrics described in "sharpness metrics". To benchmark the FAF algorithm, we also applied the conventional AF algorithm[11] using as sharpness function the 2D Brenner metric applied to the MIP along the slow axis in the full 3D reconstructed data. The speed estimation obtained from the AF was taken as the gold standard.

Clinical scanning was performed following approval from the Ethics Committee of the Klinik und Poliklinik für Dermatologie und Allergologie am Biederstein (Munich, Germany).

**Sharpness metrics**

We tested four sharpness metrics in order to identify the one optimal for the FAF algorithm and assess its performance: 2D Brenner gradient and 1D Brenner [20], mid-frequency discrete cosine transform (MDCT) [21], and maximum energy. The 2D Brenner gradient takes the reconstructed B-plane and calculates and sums the squared derivatives in the x and y directions:

$$F_{Brenner\ 2D} = \sum_{k,l}(R_{k+n,k} - R_{k,l})^2 + (R_{k,l+n} - R_{k,l})^2 \quad (2)$$

where R is the reconstructed image, k is the k-th pixel in the x direction, and l is the l-th pixel in the z direction. The reconstructed image has dimensions *k* and *l,* and *n* is a positive integer.

The 1D Brenner metric is also based on the sum of the squared derivatives, but it uses the 1D MIP of the 2D reconstruction along the z axis:

$$F_{Brenner\ 1D} = \sum_k (f_{k+n} - f_k)^2 \quad (3)$$

where *f* is the MIP of the reconstructed image *R(k,l)* along the z direction (*l*).

MDCT works as an edge detector: it convolves the reconstructed 2D B-plane with a predefined matrix and sums the resulting values:

$$F_{mdct} = \sum_{k,l}(R(k,l) * O_{MDCT})^2 \quad (4)$$

where * is the convolution operator and

$$O_{MDCT} = \begin{bmatrix} 1 & 1 & -1 & -1 \\ 1 & 1 & -1 & -1 \\ -1 & -1 & 1 & 1 \\ -1 & -1 & 1 & 1 \end{bmatrix}. \quad (5)$$

The maximum energy metric calculates the value of the pixel with the highest intensity in the reconstructed image.

## Results

Fig. 1 illustrates the geometrical considerations behind the FAF algorithm, validation of its key assumptions, and its application to simulations. Fig. 1a depicts the projection of real 3D optoacoustic sources onto a B-plane to generate the 2D distribution of virtual sources. Fig. 1b-d validate the assumptions behind the FAF algorithm. Fig. 1b shows the B-scan from a random 3D distribution of sources, while Fig. 1c shows the B-scan from the virtual sources projected onto the B-plane. Subtracting one B-scan from the other resulted in a blank image (Fig. 1d), as expected.

Fig. 1e-f show how the algorithm performed with the simulated data in Fig. 1b. Fig. 1e shows the reconstruction (speed of sound = 1,550 m/s) obtained from the B-scan of one dataset displaying the virtual sources, together with its MIP along the z axis. Fig. 1f shows the different sharpness functions; the 1D Brenner gradient led to the best cost function with a Gaussian or Lorentzian shape with the clearest maximum. Nevertheless, the 1D Brenner, 2D Brenner and MDCT metrics estimated the speed of sound for this dataset within 1 m/s of the defined value of 1,550 m/s in the forward solution.

Fig. 1g shows mean speeds of sound and corresponding standard deviations for 100 simulated datasets. The 1D Brenner, 2D Brenner and MDCT metrics correctly estimated the speed of sound with a mean value of 1,550 m/s and a standard deviation < 4 m/s.

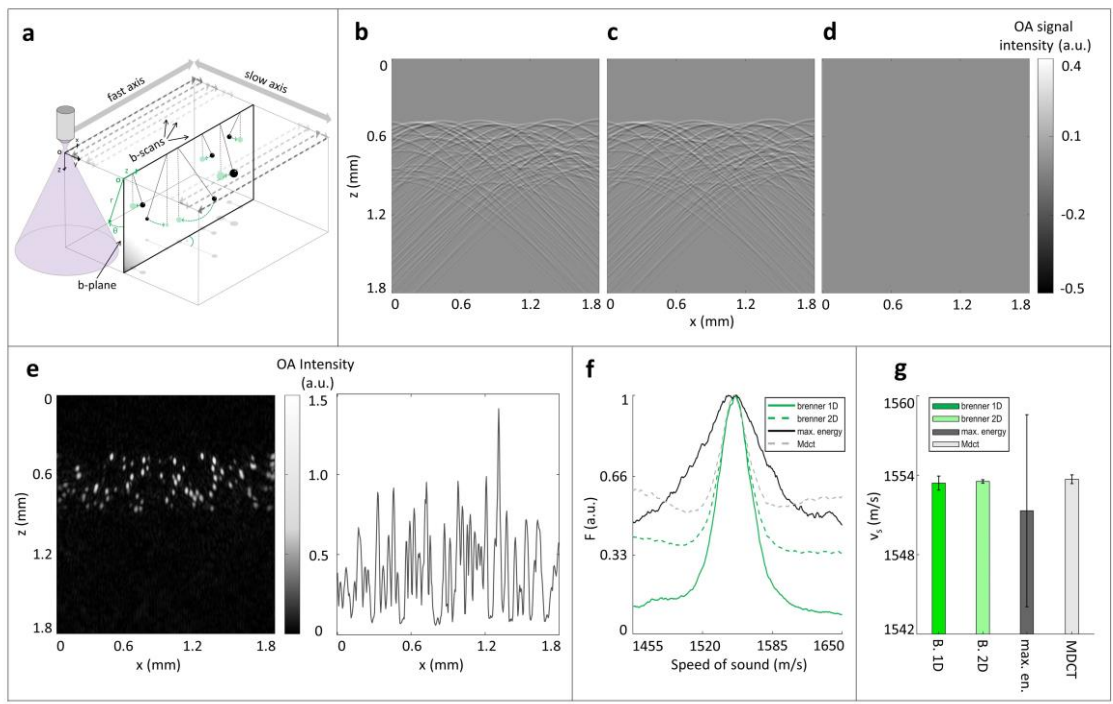

**Fig. 1. Principle of the FAF algorithm, validation of its underlying assumptions, and its application to simulated data. (a)** Scheme representing the geometrical considerations of the FAF algorithm. The transducer raster-scans (slashed lines) over the target area with its sensitivity field indicated in purple. The black bold rectangle represents a central B-plane corresponding to a central B-scan. The virtual spheres (green) used in the FAF algorithm are obtained by projecting the real spheres (black) along angle θ (dotted green arcs). **(b)** Simulated B-scan of the 3D distribution of real spherical sources. **(c)** Simulated B-scan obtained by projecting the 3D spheres onto the B-plane as described in panel (a). **(d)** Null image produced when panel (c) is subtracted from panel (b). **(e) Left:** 2D reconstruction of the B-scan from simulated data and **Right:** maximum amplitude projection along the z axis. **(f)** Calculated sharpness metrics as a function of speed of sound, based on the 2D reconstructions and 1D maximum amplitude projections corresponding to the distribution of sources in panel (b). **(g)** Mean speed of sound estimated from 100 simulated data sets by the FAF algorithm and different sharpness metrics.

Fig. 2 shows the results of applying the FAF algorithm to imaging data of a phantom (suture in water). Fig. 2a shows a central B-scan, with the "crescent-like" shape typical for an elongated suture. Fig. 2b depicts the 2D reconstruction (speed of sound = 1,495 m/s) obtained from the B-scan and its MIP along the z-axis. The different sharpness related metrics as a function of speed of sound are shown in Fig. 2c. The 1D Brenner function again showed the most prominent Gaussian or Lorentzian shape and clearest maximum. The speed of sound at the maximum, 1,495 m/s, led to a high-quality reconstructed image (Fig. 2d).

Fig. 2e shows boxplots depicting median speed of sound and associated quartiles based on 10 B-scans along the slow axis and different metrics. The 1D Brenner, maximum energy and MDCT metrics showed similarly low variation in speed of sound, while the 2D Brenner metric showed negligible variation.

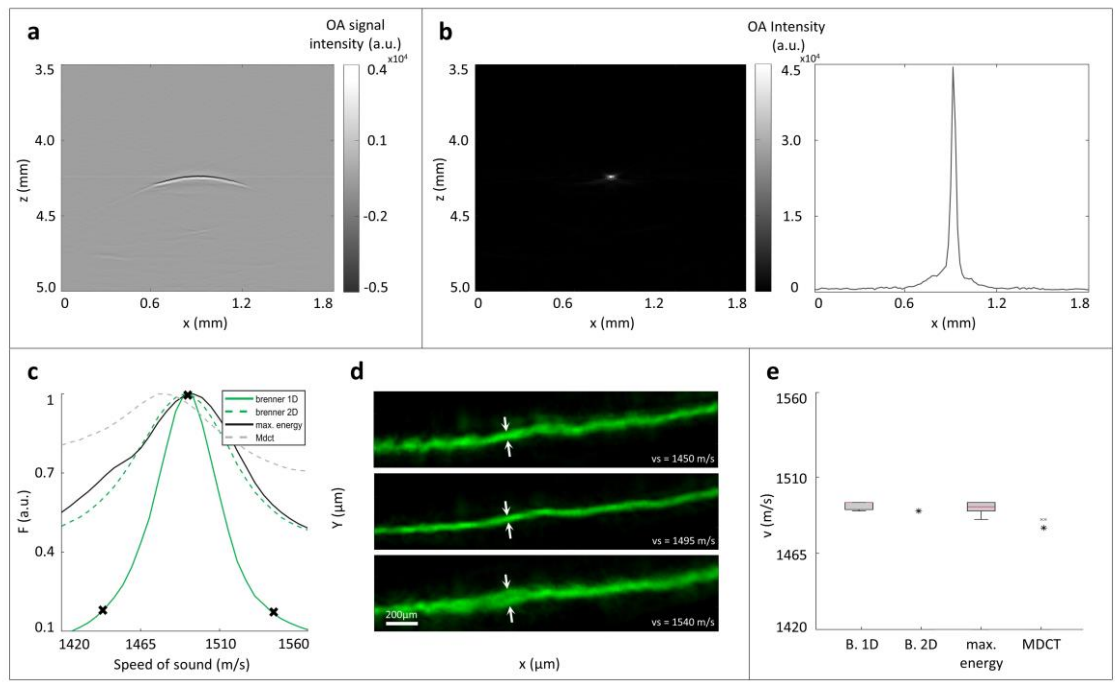

**Fig. 2. Application of the FAF algorithm to phantom data.** The phantom was a suture embedded in water (see Methods). **(a)** Sinogram corresponding to a central B-scan. **(b)** *Left:* 2D reconstruction on the B-plane corresponding to the B-scan in panel (a). *Right:* the maximum amplitude projection along the z axis. **(c)** Calculated sharpness metrics as a function of speed of sound, based on the 2D reconstructions and 1D maximum amplitude projections from the B-scan shown in panel (a). **(d)** Reconstructed images based on what the 1D Brenner gradient metric considered to be poor estimates (*top* and *bottom* rows) or a good estimate (*middle* row) of the speed of sound. See also the crosses in panel (c). **(e)** Boxplots showing the speed of sound estimated using different sharpness metrics for 10 equidistant B-scans along the slow axis.

Fig. 3 shows the application of the FAF algorithm to imaging of eczematous skin. Fig. 3a shows a central B-scan, in which the microvasculature appears as "crescent-like" shapes typical of elongated cylindrical structures, similar to the suture phantom. Fig. 3b depicts the 2D reconstruction (speed of sound = 1,515 m/s) obtained from the B-scan and its MIP along the z axis Fig. 3c shows the different sharpness metrics as a function of speed of sound. As the gold standard, we included the conventional AF algorithm, which we applied to the MIP of the reconstruction of the complete dataset (see Methods). The best metric is the Brenner function applied to the 1D maximum intensity projection, providing a healthy Gaussian/Lorentzian function with the clearest maximum. The FAF 1D Brenner estimates the same maximum as the conventional AF algorithm using the 2D Brenner method on the 3D dataset (1515 m/s) but, as expected, the computational time differs. The estimation with FAF algorithm takes in total 6.5 seconds with a reconstruction time of roughly 175 ms for one speed of sound value while the conventional autofocus algorithm applied on the whole data set needs >90 minutes to run with roughly 2.5 minutes per reconstruction. For the FAF, the value of the speed of sound corresponding to the position of the maximum lead to a high quality reconstructed images, as it can be observed in Fig. 3d. The reliability of the speed of sound estimation for 10 different B-scans along the slow axis for different metrics is shown as median values with the corresponding quartiles in Fig. 3e with the lowest variation for the 1D Brenner metric. The asterisk corresponds to speed of sound estimated by the conventional AF algorithm (Fig. 3c) which has nearly the same value as the mean value obtained from 1D Brenner function of the FAF (1518.6 m/s ± 5,5 m/s for FAF and 1515 m/s for the conventional AF algorithm).

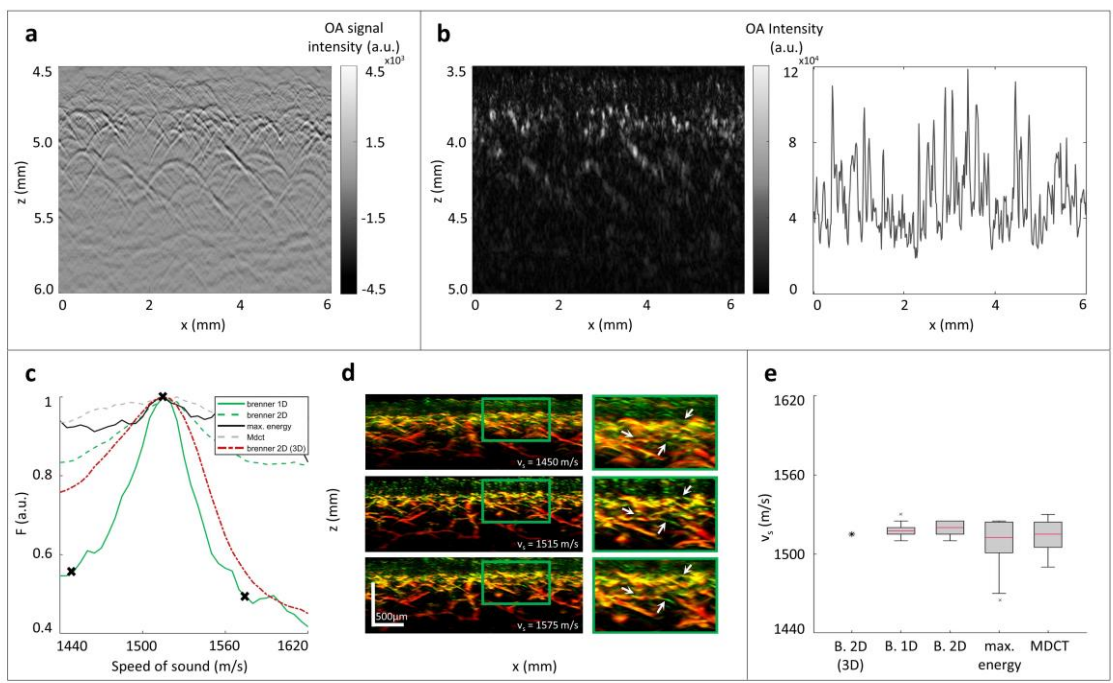

**Fig. 3. Application of the FAF algorithm to clinical data.** A region of eczematous skin was imaged (see Methods). **(a)** Sinogram corresponding to a central B-scan. **(b) Left**: 2D reconstruction on a B-plane from the B-scan in panel (a). *Right:* the MIP of the 2D reconstruction on a B-plane along the z axis. **(c)** Calculated sharpness metrics as a function of speed of sound, based on the 2D reconstructions and 1D maximum amplitude projections from the B-scan shown in panel (a). The sharpness metric obtained from the 3D reconstruction is shown in red. **(d)** Reconstructed images based on what the 1D Brenner gradient metric considered to be poor estimates (*top* and *bottom* rows) or a good estimate (*middle* row) of the speed of sound. See also the crosses in panel (c). **(e)** Boxplots showing the speed of sound estimated using different sharpness metrics for 10 equidistant B-scans along the slow axis. The asterisk shows the result for the conventional AF algorithm based on the 2D Brenner metric and the full 3D dataset.

# Discussion

In this work, we propose a FAF algorithm that overcomes the drawbacks related to the long computational times (>10 min) of its conventional counterpart. For the first time, the autofocus approach can be applied to planar geometry optoacoustic tomography systems with a large number of detectors. Such systems are rapidly progressing toward the clinic, but they require suitably fast and robust algorithms to estimate the speed of sound and other parameters. The FAF algorithm takes << 1 min to estimate the speed of sound and support the reconstruction of high-quality images.

The FAF algorithm exploits the symmetries inherent in planar geometry by operating from 2D reconstructions of B-scans as well as their 1D MIPs. The conventional AF, in contrast, must operate from 2D MIPs calculated from 3D reconstructions of complete C-scans. The dimensionality reduction in the FAF algorithm translates to substantially less computational complexity.

We found that the Brenner gradient calculated from either 2D reconstructions or their 1D MIPs generated the best cost functions for applying the FAF algorithm to real data. Similarly, the Brenner gradient appears to be optimal for the conventional AF algorithm[11]. However, the gradient is implemented with $n = 2$ in the conventional algorithm, whereas we found that the gradient with $n = 1$ performed best in the FAF algorithm (see equations 2-3).

Our results indicate that the FAF algorithm is robust to the choice of B-scan. When the algorithm was applied to 10 B-scans from clinical data, the standard deviation in the estimated speed of sound was 5.5 m/s. This variation may be reduced even more by selecting six or seven B-scans far away from one other along the entire dataset. In this case, the execution time should still be << 1 min, which is acceptable for a clinical setting and two orders of magnitude faster than the execution time of > 90 min for the conventional AF algorithm. In fact, the mean speed of sound from different B-scans of clinical data differed by only 3.6 m/s from the speed estimated by the conventional AF algorithm.

Future work will aim to adapt the FAF algorithm to calculate parameters other than speed of sound that influence image formation quality. Such parameters include aspects of detector geometry, such as focal position and numerical aperture, for which manufacturer-supplied values may not be sufficiently accurate. The adaption is straightforward as long as one aims to estimate one parameter leaving the rest fixed. To the best of our knowledge, the application of the AF algorithm on a multi-parametric space has not been reported.

A limitation of the FAF algorithm is that it assumes homogeneous speed of sound across the entire imaged volume. Major modifications will be needed to equip the algorithm to account for different speeds of sound in different regions of the target volume, such as one speed in the coupling medium and another in tissue. This situation may break the symmetries in planar geometry on which the FAF algorithm depends. Future work should compare the performance of image reconstruction with the FAF algorithm when speed of sound is assumed to be homogeneous or heterogeneous.

We expect that implementation of the FAF algorithm can bring optoacoustic tomography systems closer to the clinic in order to improve the management of dermatological, cardiovascular and metabolic diseases.

## Funding

This project has received funding from the European Union's Horizon 2020 research and innovation programme under Grant Agreement No. 687866, INNODERM (VN) and 871763, WINTHER (VN), as well as from the German Society of Dermatology "German Psoriasis Prize" (JA).